# Analysis of ultrafast magnetization switching dynamics in exchange-coupled ferromagnet-ferrimagnet heterostructures

Debanjan Polley[a,b], Jyotirmoy Chatterjee[c], Hyejin Jang[d], Jeffrey Bokor[a,b]

[a]Department of Electrical Engineering and Computer Sciences, University of California, Berkeley, California 94720 USA
[b]Lawrence Berkeley National Laboratory, 1 Cyclotron Road Berkeley, California 94720, USA
[c]IMEC, Kapeldreef 75, Leuven 3001, Belgium
[d]Materials Science and Engineering, Seoul National University, Seoul 08826, Republic of Korea

**Abstract**

Magnetization switching in ferromagnets has so far been limited to the current-induced spin-orbit-torque effects. Recent observation of helicity-independent all-optical magnetization switching (HI-AOS) in an exchange-coupled ferromagnet ferrimagnet (FM-FEM) heterostructures expanded the range and applicability of such ultrafast heat-driven magnetization switching. Here we report the element-resolved HI-AOS dynamics of such an exchange-coupled system, using a modified microscopic three-temperature model. We have studied the effect of i) the Curie temperature of the FM, ii) FEM composition, iii) the long-range Ruderman–Kittel–Kasuya–Yosida (RKKY) exchange-coupling strength, and iv) the absorbed optical energy on the element-specific time-resolved magnetization dynamics. The phase-space of magnetization illustrates how the RKKY coupling strength and the absorbed optical energy influence the switching time. Our analysis demonstrates that the threshold switching energy depends on the composition of the FEM and the switching time depends on the Curie temperature of the FM as well as RKKY coupling strength. This simulation anticipates new insights into developing faster and more energy-efficient spintronics devices.

## 1. Introduction

Ultrafast helicity-independent all-optical toggle switching (HI-AOS) has been an essential topic of research in the spintronics community for its potential applications in digital data storage [1]. It is being explored in ferrimagnet (FEM) alloys and multilayers [2, 3, 4, 5, 6, 7, 8, 9, 10, 11] over the last decade. The magnetization switches in FEMs due to the direct exchange coupling between the transition metal and rare-earth elements. However, the switching is limited within a small range of optical energy and the rare-earth element (mostly Gd) concentration and the utilization of synthetic FEMs lift these restrictions. [4] Theoretically, time-resolved and element-specific magnetization dynamics in such FEMs have been modeled by various means namely; i) atomistic Landau-Lifshitz-Gilbert equation [7, 12], ii) phenomenological mean-field theory [6, 13], and iii) a modified version of the microscopic three-temperature model

(M3TM) [14, 15, 4, 5]. Beens et al. [4, 5] have used M3TM to incorporate the direct exchange coupling between the transition metal and rare earth and described the time-resolved magnetization dynamics of FEM alloy as a function of rare earth concentration and absorbed optical energy.

On the other hand, magnetization switching in ferromagnets (FM) is generally achieved using current-induced directional spin-transfer and spin-orbit torques (different from HI-AOS) [16, 17, 18, 19]. The Landau-Lifshitz-Gilbert equation in combination with the terms for spin currents was used to analyze such a switching mechanism. Only recently, it was demonstrated that a conventional FM can also exhibit HI-AOS either due to i) the non-local ultrafast spin current [20, 21] or due to ii) the long-range Ruderman–Kittel–Kasuya–Yosida (RKKY) exchange coupling [22, 23, 24]. HI-AOS in such structures is promising as the exchange-coupled FM-FEM structure can work as a free (storage) layer in magnetic memory devices [11, 25] for efficient electrical reading due to the enhanced tunneling magneto-resistance. RKKY exchange-mediated ultrafast switching in FMs has a rich physics that has hitherto largely remained unexplored via theoretical modeling. Gorchon et al. [22] experimentally demonstrated ultrafast switching of an exchange-coupled FM layer (Co/Pt) in ∼7 ps. Recently, Chatterjee et al. [24] have obtained a significantly faster switching in ∼3.5 ps using an optimized structure and introduced a modified M3TM simulation to analyze the switching phenomenon. In these structures, RKKY-type exchange coupling [26, 27, 28] has been discussed as the primary source of angular momentum transfer channel between the FM and the FEM for obtaining ultrafast magnetization switching [22, 24].

In this paper, we discuss the theoretical modeling of the modified M3TM in greater detail and simulated the experimentally observed ultrafast magnetization switching in ∼7 ps by Gorchon et al. [22]. We have analyzed the effect of i) the Curie temperature ($T_c$), ii) the composition of the FEM, iii) RKKY coupling strength, and iv) absorbed optical energy on the ultrafast magnetization dynamics of the exchange-coupled heterostructure. The main addition in this model compared to the M3TM model, (introduced by Beens et al. [4] to study HI-AOS in FEM alloys and multilayers) is the incorporation of a long-distance RKKY exchange coupling term, which connects the magnetization dynamics of the FM and FEM layer and enables us to examine the exchange coupled heterostructure. The FM is magnetically exchange-coupled with the FEM while physically separated by a non-magnetic spacer layer (Pt) of high spin-scattering strength. The spacer layer significantly blocks any possible transmission of spin-current which can affect the magnetization dynamics [22, 24]. We observe that RKKY exchange strength, which is two orders of magnitude less than the direct exchange between the transition metal and rare earth elements of the FEM, is sufficient for switching the FM. Experimentally the Curie temperature of a Co/Pt ML system can easily be tuned by changing the thickness and the repetition of the individual layers. We detect a notable slowing of the Co/Pt switching timescale upon reducing the $T_c$ of the FM. The alloy composition of the FEM mainly affects the switching energy threshold and the dependence of the switching timescale on the RKKY exchange coupling.

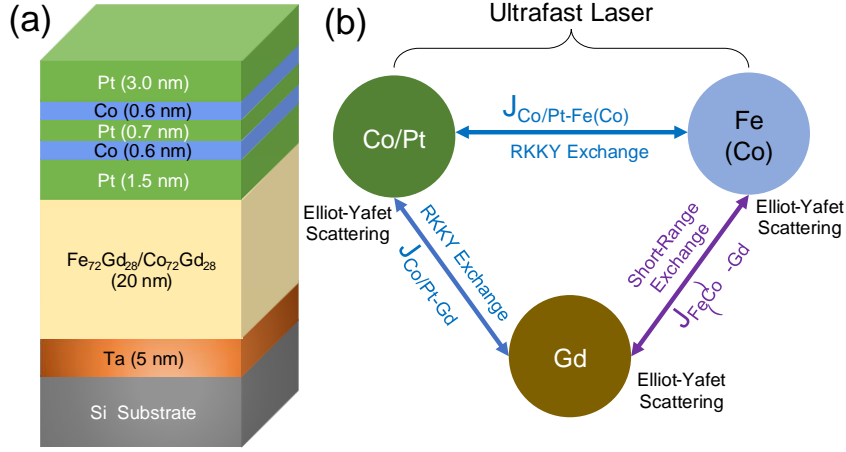

Figure 1: Schematic diagram of the simulated sample structure and (b) the different exchange interactions acting on the sub-lattices of the sample.

## 2. Sample Structure and Simulation Method

We have simulated the exchange coupled FM-FEM heterostructure (Sub/FMFEM/capping layer) experimentally studied by Gorchon et al. [22] and replaced their FEM alloy ($Fe_{65}Gd_{28}Co_7$) with $Fe_{72}Gd_{28}$ for the ease of calculation and the $T_c$ of the FM (Co/Pt) is taken to be 470 K as estimated in their experiment. We have extended the simulation to explore the effect of different FEM compositions namely; $Fe_{72}Gd_{28}$ and $Co_{72}Gd_{28}$, and henceforth, these two FEMs are denoted as FeGd and CoGd. We have simulated the effect of different $T_c$ (470 K, 530 K, and 580 K) of the Co/Pt on the magnetization dynamics. The configuration of the sample is Substrate/Ta(5 nm)/$Fe_{72}Gd_{28}$ (or $Co_{72}Gd_{28}$)(20 nm)/Pt(1.5 nm)/Co(0.6nm)/Pt(0.7 nm)/Co(0.6 nm)/Pt(3 nm), where the thicknesses of each layer are given in nanometers and schematically shown in Fig. 1a. The thickness of the Pt spacer layer, which separates the FM and FEM sub-lattices, determines the strength and sign of RKKY exchange coupling. Experimentally it was varied between 1 to 4 nm to obtain both ferromagnetic and anti-ferromagnetic coupling. The coupling strength varies non-linearly with the thickness of the Pt layer according to the characteristic oscillatory nature of the RKKY exchange interaction [26, 28] and the effect is simulated by varying the RKKY coupling within a predefined range.

The different types of exchange interactions acting on the heterostructure are schematically represented in Fig. 1b. The electron and phonon sub-systems of the FM and the two sub-lattices of the FEM have been described by a two-temperature model expressed by the equations 1 and 2. These two sub-systems (electrons and phonons) of the FM and FEM layer get excited by the ultrafast optical pulse depending on the relative optical absorption of the respective layers. First, we solve the two-temperature model to measure the temporal evolution of the electron ($T_e$) and phonon temperatures ($T_p$) for the FM and FEM using the following equations,

$$C_{e,i}\frac{\delta T_{e,i}}{\delta t} = -g_{ep,i}(T_{e,i} - T_{p,i}) + P_i; \; P_i = \frac{A_i}{\sigma\sqrt{2\pi}}e^{-\left(\frac{t}{\sigma\pi}\right)^2} \quad (1)$$

$$C_{p,i}\frac{\delta T_{p,i}}{\delta t} = g_{ep,i}(T_{e,i} - T_{p,i}) + C_{p,i}\frac{(T_{amb} - T_{p,i})}{\tau_{D,i}}; \; (i \in FM, FEM) \quad (2)$$

where $C_e$ and $C_p$ are electron and phonon specific heat, $\tau_D$ is the heat dissipation constant which represents the thermal diffusion to the substrate, $T_{amb}$ is the ambient temperature (295 K), $g_{ep}$ is the electron-phonon interaction strength and $A_i$ is the absorbed optical energy in each layer. The laser pulse is simulated as a Gaussian pulse with 100 fs full-width-half-maximum which acts as the energy source to the electronic sub-system for each layer. The laser pulse gets absorbed differently in each of the layers depending on the thickness and their optical constants, which has been evaluated by solving the transfer matrix [29] (Fig. S1 in the supplemental information) of the entire heterostructure. The complex refractive indices applied in the calculation are given in Table S1 in the supplemental information. It is assumed that the electron sub-system instantly reaches thermal equilibrium after optical excitation due to a large Coulomb interaction [30], therefore, it can be represented by specific electron temperatures, which are assumed to be homogeneous inside each layer. The phonon temperatures remain in equilibrium due to phonon-phonon interaction. All the sub-systems are initially at room temperature (295 K). The material parameters for the two-temperature model are listed in Table S2 in the supplemental information. Heat diffusion to the substrate is added to the phonon subsystem as an energy dissipation term with a 50 ps time constant ($\tau_D$) for both layers. There are two magnetic elements such as Fe(Co) and Gd comprising the FeGd (CoGd) alloy, whereas, for the Co/Pt, there is only one magnetic element. The electrons are modeled as a spinless free electron gas and the phonons are described by the Debye model according to the basic postulates of the M3TM [31]. For simplicity, it is assumed that all the spin subsystems have the same spin half quantum number ($s_i = 1/2$) [4, 5] with their respective atomic magnetic moments. We have used the following analytical expressions of exchange splitting for calculating the exchange coupling using Fermi's golden rule and the details of the magnetization dynamics are given in section S2 in the supplemental information.

$$\Delta_{Fe(Co)} = x_{Fe(Co)}\gamma_{Fe(Co)}m_{Fe(Co)} + (1 - x_{Fe(Co)})\gamma_{Fe(Co)-Gd}m_{Gd}$$

$$\Delta_{Gd} = x_{Fe(Co)}\gamma_{Gd-Fe(Co)}m_{Fe(Co)} + (1 - x_{Fe(Co)})\gamma_{Gd}m_{Gd} \quad (3)$$

$$\Delta_{CoPt} = \gamma_{CoPt}m_{CoPt} + x_{Fe(Co)}\gamma_{CoPt-Fe(Co)}m_{Fe(Co)} + (1 - x_{Fe(Co)})\gamma_{CoPt-Gd}m_{Gd}$$

where $\gamma_{ij} = zJ_{ij}\mu_j; (i,j \in Fe(Co), Gd, Co/Pt)$ can be calculated from the exchange interaction strength $J_{ij}$ and atomic magnetic moment $\mu_j$. The percentage of Fe (Co) in the FEM alloy is given by $x_{Fe}$ ($x_{Co}$); where $x_{Fe(Co)} = 0.72$. We apply the Weiss mean field theory to determine the exchange strength which has the form $J_i = \frac{3k_B T_{c,i}}{2zS_i S_{i+1}}, i = Fe(Co), Gd, CoPt$ [6]. Here z = 12 is the number of next nearest neighbors and $T_{c,i}$ is the Curie temperature of individual elements ($T_{c,Co}$ = 1388 K, $T_{c,Fe}$ = 1043 K, $T_{c,Gd}$ = 292 K, $T_{c,Co/Pt}$ = 470 K, 530 K, and 580 K). The short-range exchange interaction between Fe and Gd sub-lattices is antiferromagnetic and it has the form: $J_{Gd-Fe} = -0.348 \times J_{Fe}$ whereas, for CoGd, the coupling between Co and Gd is; $J_{Gd-Co} = -0.388 \times J_{Co}$ [3, 24]. RKKY exchange coupling strength has been varied up to ~5 % of the FM exchange i.e., $J_{RKKY} = \pm 0.05 \times J_{CoPt}$ for both the heterostructures to include the effect of varying Pt thicknesses which leads to the oscillating RKKY interaction. There are three channels for angular momentum transfer in the FEM, i) Elliott-Yafet type spin-flip scattering in Fe (Co), Gd and Co/Pt [31, 30, 32], ii) a short-range exchange coupling between Fe (Co) and Gd and iii) a long-range RKKY type exchange coupling between CoPt-Fe(Co) and CoPt-Gd layer [22]. As described in equation 3, we have used the RKKY exchange between CoPt-Fe(Co) and CoPt-Gd. However, the dominant RKKY exchange contribution comes from the interaction of Gd and Co/Pt, and by including Fe(Co) in the scheme, the dynamics are only minimally affected (Fig. S3 of supplemental information). This is because Gd has a much larger atomic magnetic moment ($\mu_{Gd} = 7.55\mu_B$, $\mu_{Fe} = 2.20\mu_B$, $\mu_{Co} = 1.72\mu_B$, and $\mu_{Co/Pt} = 1.30\mu_B$) moment compared to Fe(Co) and thereby dominates the angular momentum transfer process.

The electronic parameters of the materials in the simulation are obtained from the literature [33, 4, 34] and are given in the Table S3 of supplemental information. The time evolution of the magnetization dynamics has been calculated up to 40 ps after the optical excitation with 10 fs time resolution. The FM exchange varies linearly with $T_c$ and takes the following values; $J_{RKKY,Tc,CoPt=470K} = 1.079 \times 10^{-21}$ J, $J_{RKKY,Tc,CoPt=530K} = 1.217 \times 10^{-21}$ J and $J_{RKKY,Tc,CoPt=580K} = 1.332 \times 10^{-21}$ J. Assuming a typical atomic spacing of 0.4 nm, the RKKY exchange coupling is varied between ± 5% of the direct exchange, which results in $J_{RKKY,Tc,CoPt=470K} = \pm 0.337$ $mJ.m^{-2}$, $J_{RKKY,Tc,CoPt=530K} = \pm 0.394 mJ.m^{-2}$, and $J_{RKKY,Tc,CoPt=580K} = \pm 0.416$ $mJ.m^{-2}$ for $T_c$ = 470 K, 530 K, and 580 K respectively. S. S. P. Parkin [27] studied the exchange coupling strength of various 3d, 4d, and 5d transition metals at room temperature and found that the exchange coupling varies from 0.1 to 5 $mJ.m^{-2}$. Thus, the value of the RKKY exchange coupling strength in our simulation is well within the known range. Although the specific thickness of the Pt layer determines the sign and strength of the RKKY exchange coupling, a small thickness difference of a couple of nanometers does not change the relative absorbed energy between the FM and FEM layer by more than 4%. Therefore, for simplicity, we assume that the relative absorbed energy between the FM and FEM layer remains the same while changing the Pt spacer layer thickness (i.e. with the variation of the RKKY exchange coupling). The typical thickness variation of the Pt spacer layer (or any other spacer layer) would result in the modification of the RKKY coupling. Therefore, we have simulated the magnetization dynamics within an acceptable range of this coupling strength which is generally obtained using standard RKKY coupling layers [23, 24].

### 3. Results and Discussions

*3.1. Effect of the Curie Temperature of the Co/Pt Layer*

The phase diagrams of the magnetization as a function of absorbed optical energy and RKKY exchange coupling strength are shown in Fig. 2a-c for different values of $T_c$ (470 K, 530 K, and 580 K) of the Co/Pt layer. The color scale represents the magnetization reversal time of Co/Pt. The final magnetization state has been calculated by measuring the magnetization of each element (Fe, Gd, and Co/Pt) at 20 ps after the ultrafast laser excitation. The initial magnetization of Fe and Co/Pt is positive and it is negative for Gd before the arrival of the pump laser pulse (0 ps). Ideally, switching can be defined when the magnetization of a particular element crosses zero, however, it has been noticed that the magnetization tends to stay close to zero for some time before switching. Consequently, to avoid any ambiguity in the switching time, we have set a small non-zero magnetization value as the measure of switching [7]. The set conditions are the following in order to determine the different cases which may occur as a final magnetization state: i) relaxed if $m_{Fe}(t = 20ps) > 0.01$, $m_{Gd}(t = 20ps) < -0.01$ and $m_{CoPt}(t = 20\ ps) > 0.01$, ii) switched if $m_{Fe}(t = 20\ ps) < -0.01$, $m_{Gd}(t = 20\ ps) > 0.01$ and $m_{CoPt}(t = 20\ ps) < -0.01$, and iii) completely demagnetized if $0.01 \geq m_{Fe}(t = 20\ ps) \geq -0.01$, $0.01 \leq m_{Gd}(t = 20\ ps) \leq -0.01$, and $0.01 \geq m_{CoPt}(t = 20\ ps) \geq -0.01$.

The phonon temperature increases beyond the Curie temperature at very high optical energy, which makes the system unstable i.e. it doesn't remember its initial state, and the final state randomly switches between up and down, especially at longer timescales (> 40 ps). Such behavior has also been identified in the simulated HI-AOS properties of [Tb/Co] multilayers [35]. Hence, we have limited the calculation of the final switched state to 20 ps after the laser excitation. The FEM undergoes partial demagnetization and the FM shows either partial or full demagnetization before remagnetization at smaller absorbed optical energy. Consequently, the final state is the same as the initial state for all the elements suggesting the system is in a relaxed state irrespective of the RKKY coupling as represented by the pale-yellow region in Fig. 2a-c. In this energy range, an example of the magnetization dynamics of Fe, Gd, and Co/Pt is plotted in Fig. S4 of the supplemental information, where we find a partial demagnetization of Fe and Gd, and simultaneously Co/Pt gets fully demagnetized even though it absorbs only ~23% of the total optical energy. However, we can't switch Co/Pt due to the unavailability of the angular momentum transfer channel with the FEM (as FEM doesn't switch at smaller optical energy). With increasing optical energy, a switched region appears as depicted by the color map in combination with bright blue and magenta colors, where the magnetization reversal of both the FM and FEM sub-lattices is obtained at the same absorbed energy. The green region suggests the complete switching of the FEM layer, however, the Co/Pt layer doesn't switch. Detailed discussions about these two regions are provided later. On the other hand, with intense optical energy (dark brown region of the phase space in Fig. 2a-c), only a permanent demagnetization of all the sub-lattices is observed as the $T_p$ crosses the $T_c$ for all the magnetic components. Experimentally, in this region, the sample gets damaged and its magnetic properties can be permanently lost if we expose the sample to such high optical energy for a longer time.

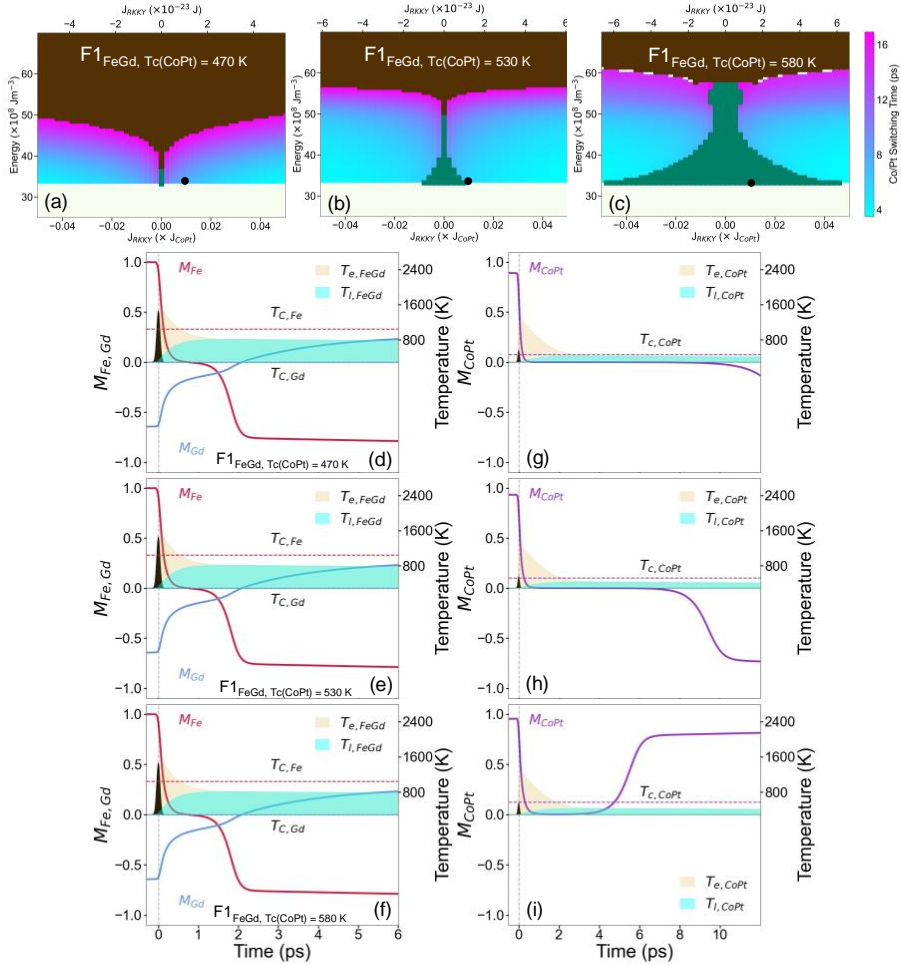

Figure 2: The phase-space of magnetization for (a) $F1_{FeGd,Tc,CoPt=470K}$, (b) $F1_{FeGd,T_{c,CoPt}=530K}$, and (c) $F1_{FeGd,T_{c,CoPt}=580K}$. The pale-yellow region shows a relaxed state, the green region depicts the switching of only FEM, the bright blue and magenta region suggests the switching of both the FM and FEM, and the dark brown region depicts complete demagnetization of the sample. The color bar of the blue and magenta region shows the switching time of Co/Pt. The time-resolved element-specific magnetization dynamics of the corresponding elements of (d-f) the FEM and (g-i) the FM sub-lattices in the FM-FEM heterostructure at the absorbed optical energy of $33.2 \times 10^8 \, J.m^{-3}$ for a specific RKKY exchange coupling ($J_{RKKY} = + 0.01 \times J_{CoPt}$). The dynamics are measured at the corresponding black dots in Fig. 2a-c. The yellow and cyan-filled region in Fig. 2d-i shows the evolution of electron and phonon temperatures of the FM and FEM layers. The red, cyan, and green dotted lines are the $T_c$s of the individual elements (Fe, Gd, and Co/Pt). The laser pulse is schematically displayed by the black-filled Gaussian pulse at time zero.

The time-resolved element-specific magnetization dynamics of the magnetic elements of the FM and FEM layers measured at the black dot on the phase space is plotted in Fig. 2d-i for absorbed optical energy of $33.2 \times 10^8 \, J.m^{-3}$ at a specific RKKY exchange coupling strength of $J_{RKKY} = + 0.01 \times J_{CoPt}$. Fe, Gd, and Co/Pt sub-lattices demagnetize at an ultrafast timescale upon ultrafast optical excitation where the demagnetization rate (at a constant fluence) is proportional to $\frac{T_c^2}{\mu}$ [31]. The transition metal (Fe) of the FEM layer, demagnetizes much faster than that of the rare earth element (Gd), and the FEM shows a transient ferromagnetic-like state between ∼0.7-1.9 ps when the magnetization of both Fe and Gd points along the same direction. As the intrinsic coupling between Fe-Gd is antiferromagnetic, Gd switches later due to angular momentum exchange with Fe. The Co/Pt layer exhibits a significantly faster demagnetization in ∼500 fs due to enhanced Elliott-Yafet scattering via spin-orbit torque as published earlier [33].

In the bright blue and magenta colored region of Fig. 2a-c, the electron temperature of Co/Pt rises faster and reaches significantly larger than its $T_c$ (can be seen from the peak of the yellow-shaded region much larger than $T_c$ in Fig. 2g-i). As a result, Co/Pt gets completely demagnetized and then its magnetization stays close to zero for a longer timescale as the electron-lattice temperature equilibrates and slowly cools down. The switched FEM kicks in to deliver enough angular momentum to ultimately switch its magnetization. In this region, the Co/Pt and the FEM switch at the same absorbed energy. The magnetization dynamics of Co/Pt show a two-step process and the switching occurs at a much longer timescale as detected in Fig. 2g-h measured at the respective black dots in Fig. 2a-b. The observation of such a two-step switching characteristic of the Co/Pt layer (exchange coupled with a FEM) has been explained as a signature of the RKKY exchange coupling mediated switching of the hot/softened Co/Pt [22, 24].

One of the main differences among the three phase-space figures in Fig. 2ac is the existence of the green region which increases with increasing $T_c$ of the Co/Pt layer, suggesting that the Co/Pt and the FEM layer doesn't switch at the same absorbed energy. FEM switches at a smaller energy (when Co/Pt remains demagnetized) and then the Co/Pt requires a larger absorbed optical energy to switch. It is relatively easy to understand that Co/Pt cannot be switched in a decoupled structure, where there is no RKKY exchange coupling with the FEM layer. Hence, we notice the green region in Fig. 2a-c, where the RKKY exchange is zero (i.e. at the middle of the x-axis). However, for larger $T_c$ (530 K, and 580 K) the green region extends towards a finite RKKY coupling strength ($-0.009 \times J_{CoPt} \leq J_{RKKY} \leq 0.009 \times J_{CoPt}$ for 530 K, and $-0.045 \times J_{CoPt} \leq J_{RKKY} \leq 0.045 \times J_{CoPt}$ for 580 K) as shown in Fig. 2b-c. Now, for a smaller $T_c$, the electron/phonon temperature requires a long time to cool below $T_c$, and the Co/Pt layer stays completely demagnetized for a longer time. The inter-sublattice (between FM and FEM) angular momentum exchange depends on the RKKY exchange strength and the available magnetization in the FEM layer, which is dominated by Gd due to its much larger atomic magnetic moment. At a longer timescale, Gd recovers a larger magnetization value as shown in Fig 2d-f, hence the effective angular momentum exchange is larger, and Co/Pt can switch at a smaller RKKY exchange strength however the switching speed gets slower. Now, with increasing $T_c$, the electron/phonon temperature gets below $T_c$ much faster as plotted in Fig. 2h, and we get Co/Pt switching at a faster timescale. The effective exchange coupling strength increases with increasing $T_c$. The value of $J_{RKKY}$ is displayed at the top x-axis of the phase space diagrams in Fig. 2a-c. Next, for an even larger $T_c$, the electron/phonon temperature cools down much more quickly to below $T_c$, as shown in Fig. 2i, when the available magnetization of Gd for the RKKY exchange is tiny. Then Co/Pt can't switch and starts to remagnetize along its initial direction. Hence, Co/Pt either needs a larger RKKY exchange coupling or a larger optical energy to switch. In case of larger RKKY exchange strength, the Co/Pt switches faster as discussed later. For larger absorbed optical energy, the electron/phonon temperature rise is higher and it takes a longer time to cool down below $T_c$. Within that time, Gd gets sufficiently magnetized in the opposite direction to provide the necessary angular momentum transfer even at a smaller RKKY exchange coupling strength. Hence, in general, Co/Pt switches at a slower speed with increasing energy as observed in Fig. S5 in the supplemental information. However, this contradicts the experimental observation by Gorchon et al. [22], where the switching speed of Co/Pt got faster with increasing optical energy. The switching dynamics also depend on the RKKY exchange coupling strength and we do notice a tiny energy range at smaller RKKY interactions, where the switching speed increases with increasing energy as plotted in Fig. S5 of the supplemental information. Another difference in the phase space among these three structures, is the larger switched region (bright blue and magenta color region), with increasing $T_c$. The Co/Pt layer can sustain larger absorbed optical energy before getting damaged, as the $T_c$ increases, hence we detect switching up to larger optical energy.

The variation of the Co/Pt switching time with the RKKY exchange strength is plotted in Fig. 3 for different $T_c$. The switching timescale slows down exponentially upon reducing the exchange coupling. As Co/Pt gets demagnetized much more quickly (plotted in Fig. 2d-i) than the FEM elements, it waits for the FEM layer (mainly Gd) to remagnetize sufficiently and provide the necessary angular momentum transfer to switch. As discussed earlier, at fixed absorbed optical energy, larger exchange coupling strength means the possibility of a larger amount of angular momentum transfer even if the Gd recovers a small magnetization in the opposite direction (in early timescale), and eventually, it translates to a faster switching. The Co/Pt layer switches ∼10 ps for $J_{RKKY} = + 0.01 \times J_{CoPt}$, for $F1_{FeGd,Tc,CoPt=470K}$, and by increasing the exchange amplitude to $J_{RKKY} = + 0.05 \times J_{CoPt}$, the switching gets faster at ∼6 ps as shown in Fig. 3a.

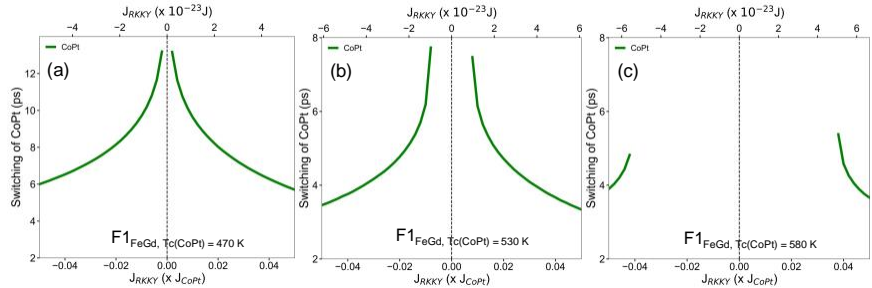

Figure 3: The variation of the switching time of Co/Pt with the RKKY exchange coupling strength for (a) $T_c = 470\ K$, (b) $T_c = 530\ K$, and (b) $T_c = 580\ K$. All the simulations are performed at absorbed optical energy of $33.2 \times 10^8\ J.m^{-3}$.

On the other hand, as we have discussed earlier, the FM with higher $T_c$ can't be switched at a smaller exchange as observed in Fig. 3b-c. A significant difference in the switching timescale (at a fixed exchange strength and absorbed optical energy) of the Co/Pt is detected for the different values of $T_c$. The switching speed of the Co/Pt layer gets faster with increasing $T_c$. At the largest RKKY exchange strength and for the fixed absorbed optical energy, with increasing $T_c$, the electron/phonon temperature reduces below $T_c$ much faster, and if sufficient angular momentum transfer channel is available, Co/Pt switches faster. Similarly, it needs a longer time, for its electron temperature to cool sufficiently below a smaller $T_c$ when the transferred angular momentum from the switched FEM sub-lattice ultimately switches its magnetization. Therefore, it is evident that the dueling time of Co/Pt in the demagnetized state (zero magnetization state) is small for larger $T_c$, which ultimately fastens the switching time. Previously we measured and simulated a differently stacked heterostructure and found the switching timescale of the Co/Pt to be ∼3.5 ps [24]. We have examined the effect of $T_c$ in that heterostructure as well and found similar dependence of the switching timescale of Co/Pt, which is plotted in Fig. S6 of supplemental information. Hence, we can also conclude that the effect of the $T_c$ on the switching time of the Co/Pt layer is universal and only minimally dependent on the sample geometry.

## 3.2. Effect of the Composition of the Ferrimagnet Layer

Next, the effect of the composition of the FEM on the magnetization switching dynamics has been explored by changing the FEM to CoGd (from FeGd). The phase space of magnetization is displayed in Fig. 4a considering CoGd as the FEM while keeping the $T_c$ of the Co/Pt fixed at 580 K. We don't see the extended green region (for a finite RKKY coupling) in Fig. 4a suggesting both FM and FEM switches simultaneously even for a larger $T_c$ (significantly different from Fig. 2c).

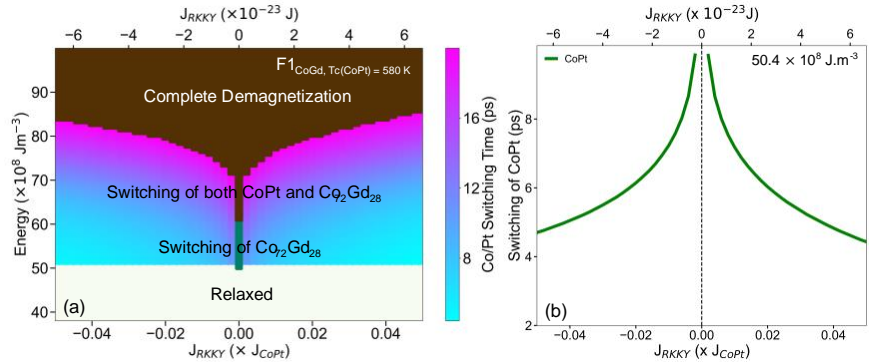

Figure 4: The phase-space of switching for $F1_{CoGd,Tc,CoPt=580K}$. The pale-yellow region indicates a relaxed state, the green region shows the switching of only FEM alloy, the bright blue and magenta region exhibits the switching of both the FM and FEM and the dark brown region shows complete demagnetization of the sample. The color bar of the blue and magenta region describes the switching time of Co/Pt. (b) Switching time of Co/Pt as a function of RKKY exchange coupling at the absorbed optical energy of $50.4 \times 10^8 \, J.m^{-3}$.

The $T_c$ of Co in the CoGd sub-lattice is much larger than that of Fe in the FeGd sub-lattice, hence larger absorbed optical energy is required to switch the CoGd FEM itself. On the other hand, the $T_c$ of the Co/Pt layer is the same as before, and ∼23% of the optical energy gets absorbed by it. Consequently, a larger optical energy is also delivered in the Co/Pt layer, and the electron/phonon temperature needs a long time to cool below its $T_c$. Within this timescale, Gd in the CoGd sub-lattice attains adequate magnetization value (discussed later), which ultimately leads to the switching of the FM and FEM layer at a finite RKKY exchange at the same absorbed optical energy. The dependence of the switching time as a function of the RKKY coupling is shown in Fig. 4b and we distinguish a similar (to Fig. 3) fastening of the Co/Pt switching time with increasing RKKY coupling strength, however, the much larger optical energy of $50.4 \times 10^8\ J.m^{-3}$ is required for switching due to the high $T_c$ of Co in the FEM layer.

The time-resolved magnetization dynamics of the FM-FEM heterostructure with different FEM sub-lattices are plotted in Fig. 5a-b at their respective threshold switching energies of $50.4 \times 10^8\ J.m^{-3}$ and $33.2 \times 10^8\ J.m^{-3}$. The red, cyan, and green lines of the figures respectively denote the magnetization dynamics of Fe (Co), Gd, and Co/Pt. The RKKY exchange strength is $J_{RKKY} = +0.05 \times J_{CoPt}$ for both systems. The demagnetization rate of Co is slightly faster than Fe (and Gd) due to its higher $T_c$ [31].

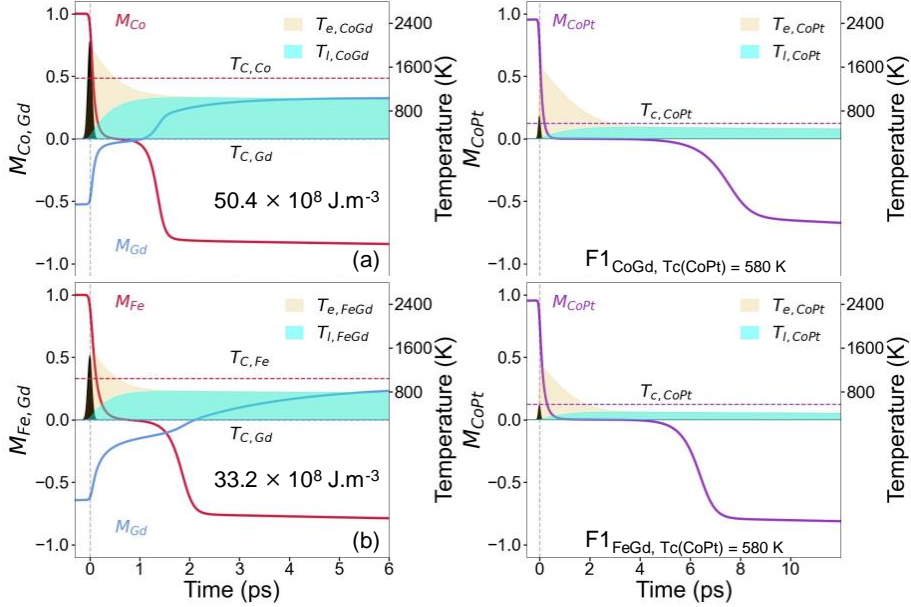

Figure 5: The time-resolved magnetization dynamics simulated for (a) $F1_{CoGd,Tc,CoPt=580K}$ and (b) $F1_{FeGd,Tc,CoPt=580K}$ measured at a particular exchange coupling strength ($J_{RKKY}= +0.05 \times J_{CoPt}$) and for respective absorbed threshold optical energy of $50.4 \times 10^8\ J.m^{-3}$ and $33.2 \times 10^8\ J.m^{-3}$. The yellow and cyan-filled region denotes the evolution of electron and phonon temperatures of the respective layers. The red, cyan, and green dotted lines are the Curie temperatures of the individual elements. The optical pulse is schematically shown by the black-filled Gaussian pulse at time zero.

On top of that, the anti-ferromagnetic intra-sublattice exchange (between rare earth and transition metal) is larger for CoGd as; $J_{Co-Gd} = -0.388 \times J_{Co} = -1.236 \times 10^{-21}$ $J$ than in FeGd as; $J_{Fe-Gd} = -0.348 \times J_{Fe} = -0.821 \times 10^{-21}$ $J$. Hence, the angular momentum transfer between Co-Gd is faster than Fe-Gd which results in a shorter transient ferromagnetic-like state of ∼0.6 ps in CoGd than ∼1.2 ps for the FeGd. Co switches at ∼0.7 ps and Gd switches at ∼1.2 ps for the CoGd alloy (whereas Fe switches ∼0.7 ps and Gd switches ∼1.9 ps for FeGd alloy). This also helps Gd to attain a significant magnetization in the opposite direction within a shorter time. Now, Co/Pt switches in ∼4.5 ps in Fig. 5a, which is slower compared to Fig. 5b (∼3.8 ps) and can be attributed to the increased threshold energy required for $F1_{CoGd,Tc,CoPt=580K}$ and faster recovery time of Gd. The maximum electron temperature rise in Co/Pt is larger at the higher threshold optical energy, and it takes longer to cool below its $T_c$, when the RKKY coupling can kick in to switch its magnetization. Therefore, the Co/Pt moments stay longer in the demagnetized state before they are flipped in the opposite direction, resulting in slightly lower and two-step switching dynamics of Co/Pt. Hence, we conclude that the effect of the composition is important for the occurrence of reversal of FM at lower RKKY exchange, and the effect of $T_c$ of plays an important role in controlling the dynamics determining the switching time of the different types of exchange coupled heterostructures.

## 4. Conclusion

We have simulated the time-resolved and element-specific magnetization dynamics in an exchange-coupled FM-FEM system using a modified M3TM. Ultrafast switching of the FM (Co/Pt) isn't possible if it is decoupled from the FEM sub-lattice, as it needs additional angular momentum transfer from the optically switched FEM sub-lattice. For a lower $T_c$, both the FM and FEM sub-lattice switches at the same threshold energy, however, with increasing $T_c$, either larger optical energy or larger RKKY coupling strength is required to switch the FM layer. Our analysis depicts the need for i) the electron/lattice temperature of the Co/Pt layer to go below $T_c$ after optical excitation and ii) the RKKY coupling strength and available magnetization from the rare-earth (Gd) element of the switched FEM alloy for the efficient angular momentum transfer to obtain ultrafast magnetization switching of the FM.

We have reproduced the experimentally observed ultrafast switching of the Co/Pt layer in at ∼7 ps by Gorchon et al. [22]. We found that the switching can be made much faster (∼3.8 ps) by increasing the $T_c$ of the FM to 580 K. We have also explored the effect of FEM composition on the magnetization dynamics. While CoGd switches at a larger threshold energy due to its high $T_c$, it displays a narrower transient ferromagnetic-like state. However, for FeGd, the transient ferromagnetic-like state lasts longer due to its weak intra-sublattice exchange interaction. On the other hand, the switching dynamics of the FM layer only get minimally affected due to the composition of the FEM layer, and that too can be attributed to the difference in the threshold optical energy.

We can conclude that the Curie temperature of the FM and RKKY exchange coupling strength are the two most important factors in tailoring the switching timescale of the FM in such exchange-coupled structures, while the composition of the FEM alloy determines the required strength of the RKKY interaction for magnetization switching. We believe our results will help in refining the current understanding of the HI-AOS phenomenon in exchange coupled structures and lead the spintronics community towards achieving a faster and more energy-efficient magnetization switching.

**Acknowledgements**

This work was primarily supported by the Director, Office of Science, Office of Basic Energy Sciences, Materials Sciences and Engineering Division, of the U.S. Department of Energy under Contract No. DE-AC02-05-CH11231 within the Nonequilibrium Magnetic Materials Program (MSMAG) (theoretical analysis). This work was also supported by ASCENT, one of the six centers in JUMP, a Semiconductor Research Corporation (SRC) program also sponsored by DARPA (instrumentation and data acquisition). We also acknowledge support from the National Science Foundation Center for Energy Efficient Electronics Science and the Berkeley Emerging Technology Research (BETR) Center (instrumentation and data acquisition).